 \documentclass[11pt]{article}

\usepackage{graphicx}
\usepackage[latin1]{inputenc}
\usepackage{amssymb,amsmath}
\usepackage{lineno}
\usepackage{natbib}
\usepackage{color}
\usepackage{amsbsy}
\usepackage{bm} 
\usepackage{sfmath}

\newcommand{\Fst}{F_{\mbox{\tiny ST}}}

\def\urltilda{\kern -.15em\lower .7ex\hbox{\~{}}\kern .04em}

\title{

{\sc A Unifying Model for the Analysis of  Phenotypic, Genetic and 
Geographic Data }
}

\author{Gilles Guillot\footnote{Corresponding author; {\tt gigu [at] imm.dtu.dk.}} 
\footnote{Informatics Department, 
Technical University of Denmark, Copenhagen, Denmark.} , 
Sabrina Renaud\footnote{Laboratoire de Biométrie et Biologie Evolutive
UMR 5558, CNRS, Université Lyon 1, Université de Lyon, 69622 Villeurbanne, France,} , 
Ronan Ledevin$^{\ddag}$,
Johan Michaux\footnote{CBGP, INRA/IRD/CIRAD/SupAgro, Montferrier-sur-Lez cedex, France.} ,  
Julien Claude\footnote{Laboratoire de Morphométrie, ISE-M, UMR5554 CNRS/UM2/IRD, Université de Montpellier II, 34095 France.}

}

\begin{document}
\bibliographystyle{sysbioAlt}

\baselineskip20pt

\maketitle
\newpage 
{\bf \centerline{Abstract}} 
{\small Recognition of evolutionary units (species, populations) requires integrating
  several kinds of data such as genetic or phenotypic markers or spatial information, in order to get a 
comprehensive view concerning  the differentiation of the units. 
We propose a statistical model with a double original advantage: 
(i) it incorporates information about the spatial distribution of the samples, with the aim to increase inference power and to relate more explicitly observed patterns to geography; and 
(ii) it allows one to analyze genetic and phenotypic data within a unified model and inference framework, 
thus opening the way to robust comparisons between markers and possibly combined analyzes. 
We show from simulated data as well are real data from the  literature that our method estimates parameters accurately  
and improves alternative approaches in many situations. 
The interest of this method is exemplified using an intricate case of inter- and intra-species differentiation based on an original data-set of 
georeferenced genetic and morphometric markers obtained on {\em Myodes  } voles from Sweden. 
A computer program is made available as an extension of the R package Geneland.
}

{\bf \centerline{Keywords}} 
{\small Clustering, spatial data,  bio-geography, Bayesian model, 
Markov chain Monte Carlo, R package, morphometrics,  molecular markers, {\em Myodes}.}

\newpage 

Species delimitation are of interest in conservation biology (identification and management of  endangered species), 
epidemiology (detection of new pathogens) but also from a purely cognitive point of view to describe, quantify 
and understand mechanisms of speciation. 
Methodological advances in evolutionary biology have led to methods for species 
identification solely based on the variation of key genetic markers \citep[e.g. DNA barcoding,][]{Luo11}. 
Limits of these single-marker approaches are more and more evidenced by conflicts between different genes in a multi-marker approach 
\citep{Rodriguez10,Turmelle11} or between genetic and phenotypic markers \citep{Nesi11}.
In this context of species or population identification, phenotypic data still emerge of interest together with genetic markers. \\
Phenotypic data such as size and/or shape of morphological structures are the product of numerous interacting nuclear genes 
\citep{Klingenberg01} and as such can provide a global estimate of the divergence between units. 
Furthermore, by being the target of the screening by selection, morphological variation can provide precious insights 
on the selection pattern contributing to shape the units. In the case of fossil lineages, it may even be the only 
information available to identify evolutionary and systematic units \citep{Neraudeau11,Girard11}. 

A rich toolbox is available to tackle these questions. 
Many  methods work as partition clustering, and aim at defining how many groups are represented in a sample of individuals, 
and assign these individuals to these groups following some optimality principles. These methods were initially  developed to deal with 
continuous quantitative measurements. These classical clustering methods have been implemented in programs 
such as {\sc Emmix} \citep{Mclachlan99} or {\sc Mclust} \citep{Fraley99}  or {\sc Mixmod} \citep{Biernacki06}. 
The methods above did not  received a strong interest in Systematics until recent  Population Genetics extensions to deal with molecular 
data such as the widely used 
computer program {\sc Structure} \citep{Pritchard00} and related work \citep[reviewed e.g. by][]{Excoffier06}. 
More recently, \citet{Hausdorf10} and \citet{Yang10} developed methods 
for delimiting species based on multi-locus data. While the approach of  \citet{Hausdorf10} method hinges on Gaussian clustering, 
the method  of \citet{Yang10} is based on the coalescent and makes use of a user-specified guide tree. 
Methods for genetic data have been also extended to incorporate information about the spatial location of each sample 
- an information rarely used although commonly available in data analysis in evolutionary biology - 
with the aim of increasing power of inferences and of relating more explicitly observed patterns to geography \citep{Guillot05a,Guillot09g}.   

These tools have been developed by different communities (evolutionists, population geneticists, statisticians). 
Therefore, one still lacks a unified framework, and this constitutes a major drawback for combining various kinds of data. 
This is especially true for morphological markers that did not received as much attention as genetic markers for recognizing populations and species. 
There are therefore a few major gaps in the toolbox available to identify evolutionary units, namely there is to date: 
no method to analyze genetic data and phenotypic data  under the same general paradigm (model and inference framework), 
and no method  to incorporate  spatial information in such  phenotypic/genetic analysis.

The goal of the present paper is to fill these gaps.
We  propose a  model to deal in an integrated way with georeferenced phenotypic and genetic data and we provide 
a computer program freely available that implements this model and should ease data analysis in many respects. 
Given the complexity of the modeling and inferential task, our method is not based on an explicit evolutionary model 
(for example based on the coalescent) but on a statistical model. This model is a parametrization which is general 
enough to capture some essential features in the data variation, but also simple enough 
to be subject to a rigorous and accurate inference method. Briefly, our model assumes the existence of several 
clusters which display some kind of homogeneity. This model mimics more or less what would be expected from a population: 
homogeneity in terms of genetic and phenotypic variation and some geographical continuity. 
The existence of homogeneous clusters corresponds to the fact that some individuals have shared some aspects of their 
recent ecological or evolutionary history. This shared history is summarized by cluster-specific parameters which are 
allele frequencies and  means and variances of phenotypic traits. 
Because it is not based on an explicit evolutionary model, it does not require prior information 
(as for instance a guide tree in the case of Yang and Rannala's method).
The statistical challenge in this context is to estimate 
the number of clusters and these cluster-specific parameters. \\
This article is organized as follows. First we provide a description of the model and inference machinery. 
Next we illustrate our method and test its accuracy on a large set of simulated data as well as on two published real data-sets. 
Then we implement our method on an original data-set of georeferenced 
genetic and morphometric markers to decipher the complex inter-and intra-specific structure of red-backed and bank voles 
{\em Myodes rutilus} and {\em M. glareolus} in Sweden. We conclude by discussing potential applications in a more general context.

\clearpage 
\section*{\sc Method}

\subsection*{Overview} 
We assume that we have a data-set consisting of  $n$ individuals sampled at sites ${\bf s}=(s_i)_{i=1,...,n}$ 
(where $s_i$ is the two-dimensional spatial coordinate of individual $i$),  
observed at some phenotypic variables denoted  ${\bf y}=(y_{ij})_{\substack{\;i=1,...,n \\j=1,...,q}}$ 
and/or some genetic markers denoted ${\bf z}=(z_{ij})_{\substack{\;i=1,...,n \\j=1,...,l}}$. 
Our approach is able to deal with any combination of  phenotypic and genetic data, 
including situations where only phenotypic or only genetic data are available 
and situations when each individual is observed through its own combination of phenotypic and genetic markers. 
As it will be shown below, our approach also encompasses the case where sampling locations are missing (or considered to be 
irrelevant). The only constraint that we impose at this stage is that if spatial coordinates are used, 
they must be available for all individuals.
We assume that each  individual sampled belongs to one of $K$ different clusters and that variation in the data 
can be captured by cluster-specific location and scale parameters.

\subsection*{Prior and Likelihood Model for Phenotypic Variables}
Denoting by $p_i$ the cluster membership of individual $i$ ($p_i \in \{1,...,K\}$), 
we assume that conditionally on $p_i=k$, $y_{ij}$ is drawn from a parametric distribution with cluster-specific parameters. 
Independence is assumed within and across clusters  conditionally on cluster membership. This means in particular that 
there is no residual dependence between variables not captured by cluster memberships. 
Implications of this assumption are discussed later.
Although most of the analysis that follows would be valid for all families of continuous distribution, 
we assume in the following that the $y$ values arise from  a  normal distribution. 
Each cluster is  therefore characterized by a mean $\mu_{kj}$ and a variance $\sigma^2_{kj}$
and our model is a mixture of multivariate independent normal distributions \citep{FruhwirthSchnatter2006}.
Following a common practice in Bayesian analysis \citep{Gelman04}, we use the natural conjugate prior family on 
 $(\mu_{kj},1/\sigma^2_{kj})$ for each cluster $k$ and variable $j$.
Namely, we  assume that the precision $1/\sigma^2_{kj}$ (i.e.  inverse variance)  follows  a Gamma distribution ${\cal G}(\alpha_{},\beta_{})$
($\alpha_{}$ shape, $\beta_{}$ rate parameter) and that conditionally on $\sigma_{kj}$, the mean $\mu_{kj} $ has a normal distribution  with 
mean $\xi_{}$ and variance $\sigma^2_{kj} / \kappa_{}$. 
In the specification above, $\alpha_{},\beta_{}, \xi_{}$ and $\kappa_{}$ are hyper-parameters. Details about their choice 
are  discussed in the appendix and in the supplementary material.

\subsection*{Prior and Likelihood Model for Genetic Data}
We assume here  a mixture  of multinomial distributions. This is the model previously introduced by \citet{Pritchard00} 
to model individuals with pure ancestries. 
Denoting frequency 
of allele $a$ at locus $l$ in cluster $k$ by $f_{kla}$, for diploid genotype data we assume that
\begin{eqnarray}
\pi(z_{ij} =\{a,b\}| p_i = k) & = & 2 f_{kla} f_{klb} \;\; \mbox{ whenever } \;\; a \neq b \\
\mbox{ and }\pi(z_{ij} =\{a,a\}| p_i = k) & = & f_{kla}^2.
\end{eqnarray}
While for haploid data, we have 
\begin{eqnarray}
\pi(z_{ij} = a | p_i = k) & = & f_{kla} 
\end{eqnarray}
We also deal with dominant markers for diploid organisms with a modified likelihood \citep[see][for details]{Guillot10b,Guillot11b}.
We assume independence of the various loci within and across clusters conditionally on cluster memberships. 
In particular, as with all other population genetic clustering models 
(including {\sc Structure} ), we do not attempt to model background linkage disequilibrium (LD). 
Therefore, our model can handle non-recombining DNA sequences (such as data obtained from mitochondrial DNA, 
Y chromosomes or tightly linked autosomal nuclear markers) 
provided data are reformatted in such a way that the various haplotypes are recoded as alleles of a single locus, but see also discussion.
We assume that allele frequencies $f_{kl.}$ have a Dirichlet distribution. Independence of the vectors 
$f_{kl.}$ is assumed across loci. Regarding the dependence structure across clusters, 
we consider either independence (referred to as  Uncorrelated Frequency Model or UFM) 
or an alternative model (referred to as  Correlated Frequency Model or CFM) 
introduced by \citet{Balding95,Balding97}.
In this second model, allele frequencies also  follow a Dirichlet distribution but now depending on some cluster-specific drift parameters. 
In this model, $f_{kl.}$ are assumed to follow a Dirichlet distribution ${\cal D} (\tilde{f}_{la} (1-d_k)/d_k,...,\tilde{f}_{lA}(1-d_k)/d_k)$ 
where  $d_k$s parametrize the speed of divergence of the various clusters and the  $\tilde{f}_{la}$s represent the allele frequency 
in an hypothetical ancestral population. This model can be viewed as  a heuristic and computationally convenient approximation 
of a scenario in which present time clusters result from the split of an ancestral cluster some generations ago.
It is also a Bayesian way of introducing correlation between clusters at the allele frequency level and hence to infer subtle differentiations 
 that would have been missed by a model assuming independence of allele frequencies across clusters \citep{Falush03,Guillot08b,Siren11} . 

\subsection*{Prior Models for Cluster Membership}
\subsubsection*{Spatial model}
\begin{figure}
\begin{tabular}{ccc}
\hspace{-1cm}\includegraphics[width=6cm]{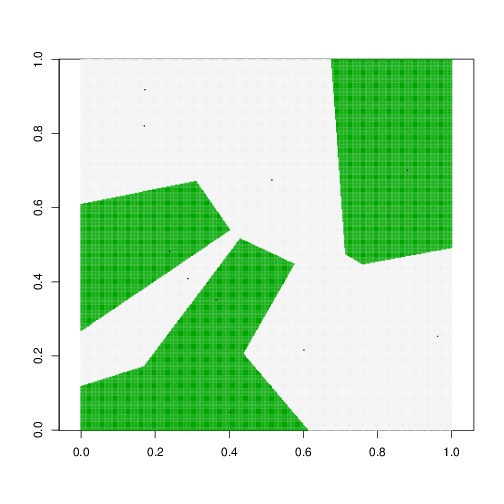} & \hspace{-1cm} \includegraphics[width=6cm]{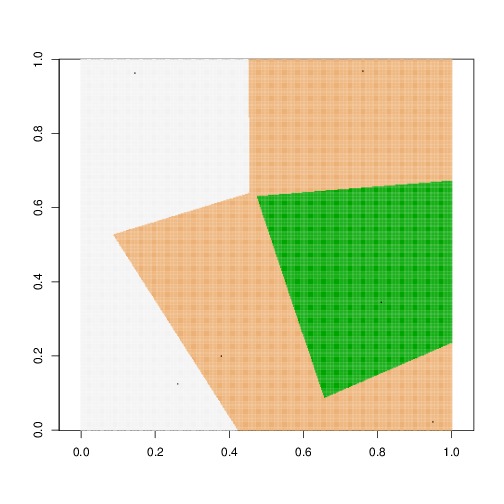} & \hspace{-1cm} \includegraphics[width=6cm]{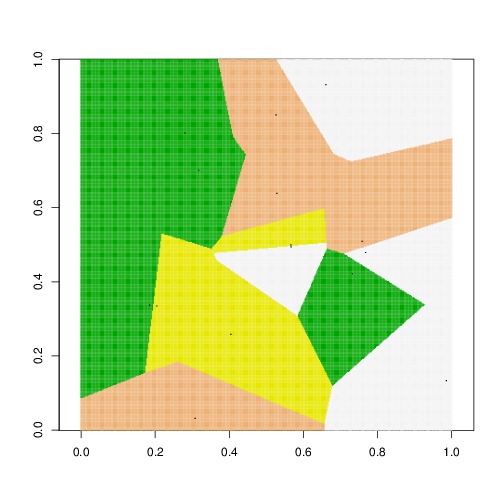} 
\end{tabular}
\caption{Examples of spatial clusters simulated from our prior model. The square represents the geographic study area. 
Membership of a geographical site to one of the $K$ clusters is coded by a color. 
From left to right: $K=2,3$ and $4$. A given clustering depends on $K$, and on the number, locations and colors (cluster memberships) 
of each polygon. If the prior placed on the number of polygons tends to favors low values, then each cluster tends to be made of one or only a few 
large areas. This is in sharp contrast with non-spatial Bayesian models which typically assume that clusterings 
with highly fragmented cluster areas are not unlikely.}\label{fig:sim_tess}
\end{figure}

We consider a statistical model known as colored Poisson-Voronoi tessellation.
Loosely speaking, this model assumes that each cluster area in the geographic domain can be approximated by the union of a few polygons.
Most of the modeling ideas can be grasped from the examples shown in figure \ref{fig:sim_tess}. 
The polygons are assumed to be centered 
around some points that are generated by a homogeneous Poisson process (i.e. points located completely at random in the geographic domain).
Formally,  we denote by $(u_1,...,u_m)$  
the  realization of this Poisson process. These points in $\mathbb{R}^2$ induce a Voronoi tessellation into $m$ subsets 
 $\Delta_1,...,\Delta_m$ . The Voronoi tile associated with point $u_i$ is defined as
$\Delta_i = \{s \in \mathbb{R}^2, dist(s,u_i) < \mbox{dist}(s,u_j) \forall j \neq i \}.$
Each tile receives a cluster membership $c_i$ (coded graphically as a color hence the terminology) at random  sampled independently from 
a uniform distribution on $\{1,...,K\}$.
Denoting by $D_k$ the union of tiles with color $k$, 
the set $(D_1,...,D_K)$ defines a tessellation in $K$ subsets. This model is controlled by the intensity of the Poisson process 
 $\lambda$  (the average number of points per unit area) 
and the number of clusters $K$.
We place a uniform prior on $[0,\lambda_{max}]$  and on $\{0,...,K_{max}\}$ respectively. 
This model is a flexible tool widely used in 
engineering to fit arbitrary shapes  in a non-parametric way \citep{Moller09}. 
It offers a good trade-off between model complexity, realism and computational efficiency. 
It is presumably most useful in situations of incipient allopatric speciation 
but examples of applications in other contexts can be found e.g. in the studies of \citet{Coulon06,Fontaine07,Wasser07,Hannelius08,Joseph08,Sacks08,Galarza09,Beadell10}. 
See also \citet{Guillot09g} for review and additional references. 
Lastly, we note that our approach relates to that of \citet{Hausdorf03} who propose a test for clustering of areas of distribution. 
However,  rather than testing clusteredness, our approach estimates these areas of distribution. 
To do that, we assume some clusteredness but without making strong assumptions about its intensity.

\subsubsection*{Non-spatial model}
If spatial coordinates are not available or thought to be irrelevant to the species at the spatial scale considered, then a non-spatial 
model can be used. The non-spatial modeling option considered here does not require to introduce any auxiliary point process as above 
but for the sake of consistency, we use the same setting as in the paragraph above.
We set $m=n$ and impose $(u_1,...,u_n) = (s_1,...,s_n)$. 
Here the  $s_i\;$s are some known spatial coordinates or dummy points if this piece of information is missing. 
This model does not impose any spatial structure and corresponds to the model implemented in most non-spatial cluster programs,  
including the genetic clustering programs {\sc Baps} \citep{Corander03,Corander04} and {\sc Structure} 
(with the exception of the latest model presented by \citet{Hubisz09}.

\subsection*{Summary of Proposed Model}

The parameters in our model are as follows: 
number of clusters $K$, 
rate of Poisson process $\lambda$, 
number of events (points) of the Poisson process $m$, 
events of Poisson process ${\bf u}=(u_1,...,u_m)$, 
color of tiles (i.e. cluster membership of spatial partitioning sub-domains)  ${\bf c}=(c_1,...,c_m)$, 
allele frequencies ${\bf f}=(f_{kla})$ (frequency of allele $a$ at locus $l$ in cluster $k$), 
genetic drift parameters ${\bf d}=(d_1,...,d_K)$, 
allele frequencies in the ancestral population ${\bf \tilde{f}}=(\tilde{f}_{la})$,
expectations of phenotypic variables ${\boldsymbol \mu}=(\mu_{kj})$, 
standard deviations  of phenotypic variables ${\boldsymbol \sigma}=(\sigma_{kj})$ 
(note that ${\boldsymbol \sigma}$ is not a variance-covariance matrix (the phenotypic variables are assumed to be independent) 
but rather a set of scalar variances stored in a two-dimensional  array.
On top of this, we place a uniform prior on $[0,\lambda_{max}]$ on $\lambda$, 
a uniform prior on $\{0,...,K_{max}\}$ on $K$, a Beta $\mbox{B}(\delta_k,\delta_k))$ prior on $d_k$ 
and a Gamma distribution ${\cal G}(g,h)$ on $\beta$.

The vector of unknown parameters is therefore 
${\boldsymbol \theta}=(K,\lambda,m,{\bf u},{\bf c},{\bf f},{\bf \tilde{f}},{\bf d},
{\boldsymbol \mu},{\boldsymbol \sigma},{\boldsymbol \beta})$. 
We also denote by ${\boldsymbol \theta}_S=(\lambda,m,{\bf u},{\bf c})$, ${\boldsymbol \theta}_G=({\bf f},{\bf \tilde{f}},{\bf d})$ 
and ${\boldsymbol \theta}_P= ({\boldsymbol \mu},{\boldsymbol \sigma},{\boldsymbol \beta})$ 
the parameters of the spatial, genetic and phenotypic parts of the model respectively.

\begin{figure}[h]
\vspace{-2cm}
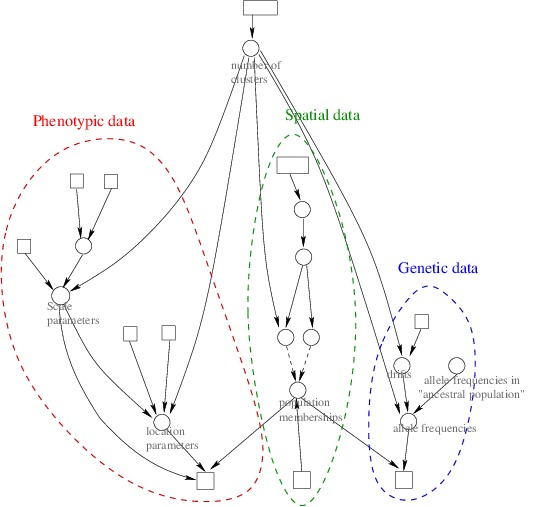
\caption{Graph of proposed model. Continuous black lines represent stochastic dependencies, dashed black lines represent 
deterministic dependencies. Boxes enclose data or fixed hyper-parameters, circles enclose inferred parameters. 
Bold symbols refer to vector parameters. 
The red, green and blue dashed lines enclose parameters relative to the phenotypic, 
geographic and genetic parts 
of the model respectively. The parameters of interest to biologists are the number of clusters $K$, 
the vector ${\bf p}$ which encode the cluster memberships,
and possibly  allele frequencies ${\bf f}$, mean phenotypic values $\bm{ \mu}$, phenotypic variance  $\bm{ \sigma^2}$ 
which quantify the genetic and phenotypic 
divergence between and within clusters. Other parameters can be viewed mostly as nuisance parameters.}\label{fig:dag}
\end{figure}

The hierarchical structure of the model is summarized on the graph shown in figure \ref{fig:dag}. 
There are three blocks of parameters relative  to the genetic, phenotypic and geographic component of the model. 
Information propagates from data to higher levels of the model across the various nodes of the graph through probabilistic relationships 
specified between neighboring nodes. 

The structure of the global model can be summarized by the joint distribution of ${\boldsymbol  \theta}$ and $({\bf y},{\bf z })$. 
By the conditional independence assumptions, we get
\begin{eqnarray}
 \pi({\boldsymbol  \theta, \bf y},{\bf z })  &  = & \pi({\boldsymbol \theta}) \pi({\bf y},{\bf z} | {\boldsymbol \theta}) \nonumber \\
 &  = & \pi({\boldsymbol \theta}) \pi({\bf y} |{\boldsymbol \theta}) \pi({\bf z} |{\boldsymbol \theta})     \nonumber\\
 &  = & \pi({\boldsymbol \theta}) \pi({\bf y} |{\boldsymbol \theta_P}) \pi({\bf z }|{\boldsymbol \theta_G}) 
\end{eqnarray}
Each genetic or phenotypic marker brings one factor in the likelihood. 
Whether the clustering is driven by the genetic or the phenotypic data depends on the respective differentiation and on the number of markers 
of each kind.

\subsection*{Estimation of Parameters}
\subsubsection*{Bayesian estimation and Markov chain Monte Carlo inference}
We are interested in the 
posterior distribution $\pi({\boldsymbol \theta} | {\bf y},{\bf z})$. 
Note that this notation does not refer explicitly to the sample locations because, unlike genetic markers and  phenotypic variables, 
locations are not considered as random quantities in our model.  The model does in fact implicitly account for spatial information.
The distribution $\pi({\boldsymbol \theta} | {\bf y},{\bf z})$ is defined on a high dimensional space and deriving properties analytically 
about this distribution is out of reach. 
We implement a Markov chain Monte Carlo strategy. 
This amounts to generating a sample of $N$ correlated replicates $({\boldsymbol \theta}_1,...,{\boldsymbol \theta}_N)$
from the posterior distribution $\pi({\boldsymbol \theta} | {\bf y},{\bf z})$. 
The initial state ${\boldsymbol \theta}_1$ is simulated at random from a distribution that does not matter in principle,  
a fact that has to be checked in practice by convergence monitoring tools \citep{Gilks96,Robert04}. 
We always sample  ${\boldsymbol \theta}_1$ from the prior and we check that starting from various random states  
does not affect the overall result provided a suitable number of burn-in iterations are discarded.
In analyzes reported below, the order of magnitude of $N$ was 50000-100000 iterations with 20000 burn-in iterations. 
See appendix for detail on the MCMC algorithm. 

\subsubsection*{Estimation of the number of clusters}
Each simulated state ${\boldsymbol \theta}_i$ includes a simulated number of clusters $K_i$. 
The number of clusters is estimated as the most frequent value among the $N$ simulated values  $K_1,...,K_N$ and we denote it by $\hat{K}$. 

\subsubsection*{Estimating cluster memberships}
A model assuming that individuals $i$ and $j$ belong respectively to clusters 1 and 2  characterized by a mean phenotypic trait equal to 5 and 7
is essentially the same as 
a model assuming that individuals $i$ and $j$ belong respectively  to clusters 2 and 1  characterized by a mean phenotypic trait equal to 7 and 5.
This trivial fact is due to the invariance of the likelihood under permutation of cluster labels and 
 brings up a number of computational difficulties in the post-processing 
of MCMC algorithm outputs known as the label switching issue \citep{Stephens97}. 
In particular, it does not make sense to average values across the MCMC iterations. 
To deal with this, we implement the strategy described by  \cite{Marin05} and \citet{Guillot08b}. 
We consider the set of simulated   ${\boldsymbol \theta}$ values restricted to the set of states such that $K=\hat{K}$. 
Then working on this restricted set, we relabel each state in such a way that they ``best look like'' the modal state of the posterior distribution.
Cluster memberships of each individual are estimated as the modal value in this relabeled sample. 
Then we estimate all cluster-specific parameters (mean phenotypic values and allele frequencies) by taking 
the average simulated value over the relabeled sample.

\section*{\sc Analysis of Simulated Data}

We investigate here two new aspects of the model, namely its ability to cluster phenotypic data only and  phenotypic and genetic data jointly 
together with some spatial information. 

\subsection*{Inference from Phenotypic Data Only}

\begin{figure}[h]
\begin{tabular}{cl}
\hspace{-1cm}\includegraphics[width=9cm]{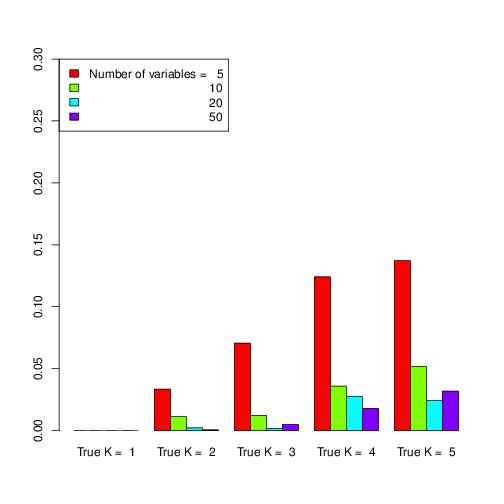} & %
\hspace{-1.cm}\includegraphics[width=9cm]{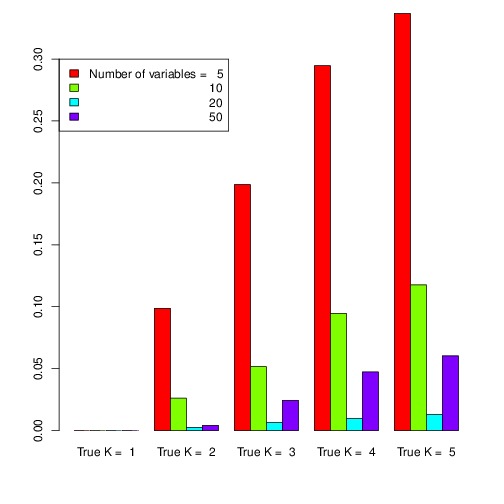}\\
\small (a) Our method &  \small (b) Mclust
\end{tabular}
\caption{Classification error from simulated data. The variable plotted on the y-axis is the proportion of misclassified individuals
(after correction for potential label switching issues). Each bar is obtained as an average over 500 data-sets consisting of $n=200$ individuals. 
Both methods are excellent at avoiding false positives (i.e. reporting $\hat{K}=1$ when $K$=1) and have 
a clear ability to reduce the error rate when the number of variables increases. They seem to lose accuracy  in the same fashion when they 
are given an increasingly  difficult problem (i.e. when the true $K$ increases) and have difficulty fully exploiting all of the available information 
when the number of variables is large (cf. loss of accuracy for 50 variables compared to 20 variables).
In the overall, under this type of simulated data, our method is typically twice as accurate as the competing method.
}\label{fig:res_error_barplot_nindiv=200_spatial=TRUE}
\end{figure}

In this section, we present new results on the model for phenotypic data and focus on the spatial model option.
We carried out simulations from our prior model and performed inferences as described in section ``Estimation of parameters'' above. 
We produced data-sets consisting of $n=200$ individuals with $q=5,10,20$ and $50$ phenotypic variables. 
For each value of $q$, we produced 500 data-sets 
with a uniform prior ${\cal U}(\{1,...,5\})$ on $K$. 
In real-life, the range of value of the putative true $K$ is largely unknown. 
To be as close as possible to this situation, 
we carried out inference under a  uniform ${\cal U}(\{1,...,10\})$ prior for $K$. 
We assessed the accuracy of inferences by computing the classification error which is displayed in figure 
\ref{fig:res_error_barplot_nindiv=200_spatial=TRUE}. Further details are provided in Supporting Material. 

We also wished to assess how our method performs compared to other computer programs implementing state-of-the-art methods.
We therefore considered the  R package {\sc Mclust} \citep{Banfield93,Fraley99} which is one of the most 
widely used and arguably  most advanced  program to perform clustering. 
This program implements inference for Gaussian mixtures and as such  deals solely with continuous quantitative data. 
It implements a non-spatial algorithm and in its default setting performs inference by likelihood maximization 
via the Expectation Maximization (EM) algorithm. It implements a wide class of sub-models regarding the covariance structure of the data. 
In its default option (which we used) it performs model selection (covariance structure and number of clusters) 
by optimizing a  Bayesian Information Criterion (BIC). 
We set the maximum number clusters to the $K_{max}=10$, i.e. to the same value as in analyzes with our method.

We stress here that the goal of this experiment is not to rank our method and  {\sc Mclust} as the two methods/programs differ 
in many important respects. 
They differ regarding 
the type of data handled ({\sc Mclust} is not aimed at genetic data  and does not implement any spatial model) and the breadth of covariance 
structure considered (our approach assumes conditional independence while {\sc Mclust} considers in excess of ten types of covariance structures).
It would be  therefore difficult to design an efficient and fair comparison. 
Results are mostly given here to support the claim that our method compares with  state-of-the-art methods and 
to assess the magnitude of improvement brought by the use of a spatial 
model in a best-case scenario when data are spatially structured (see also discussion).
Most of the numerical results are summarized in figure \ref{fig:res_error_barplot_nindiv=200_spatial=TRUE}.

\begin{figure}[h]
\begin{tabular}{cc}
\hspace{-0cm}\includegraphics[width=8cm]{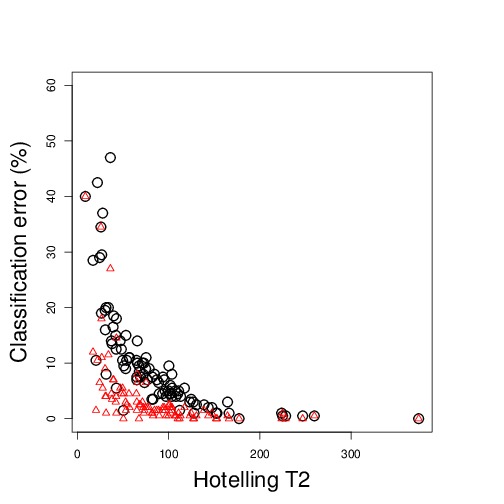} & 
\hspace{-0cm}\includegraphics[width=8cm]{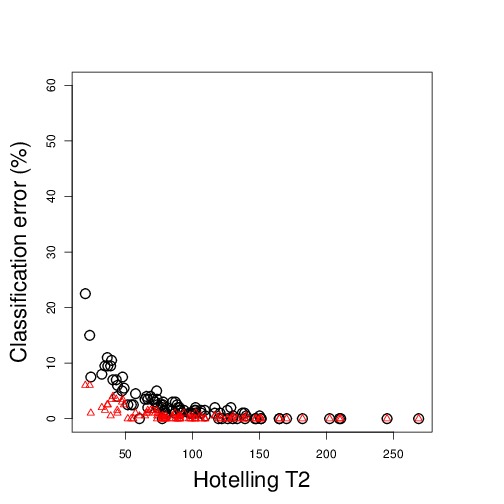} \\
$q=5$ & $q=10$ \\
\hspace{-0cm}\includegraphics[width=8cm]{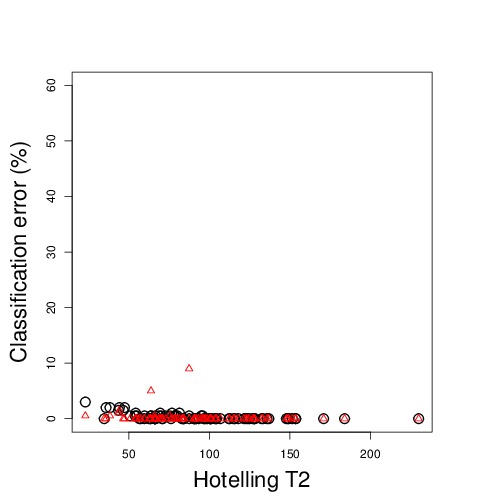} & 
\hspace{-0cm}\includegraphics[width=8cm]{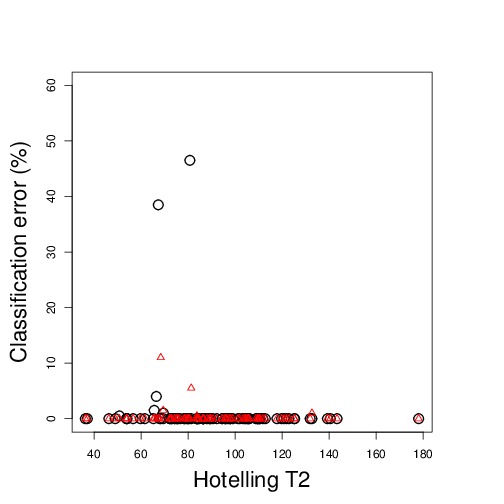} \\
$q=20$ & $q=50$ \\
\end{tabular}
\caption{Classification error for simulated data-sets consisting of $K=2$ clusters as a function of the phenotypic differentiation between the clusters.
The variable plotted on the y-axis is the proportion of misclassified individuals
(after correction for potential label switching issues). 
The variable plotted on the x-axis is the Hotelling T statistic and  assesses the magnitude of the phenotypic differentiation. 
Our method: red triangles ({\color{red} \footnotesize  $\boldsymbol  \triangle$}), 
{\sc Mclust}: black circles ({$\boldsymbol \circ$}). }\label{fig:Hotelling_nindiv=200_spatial=TRUE}
\end{figure}

To understand better how the method behaves as a function of the pairwise phenotypic differentiation between clusters, 
we also report the classification error as a function of the $T^2$ statistic in a Hotelling T test \citep{Anderson84} on 
figure \ref {fig:Hotelling_nindiv=200_spatial=TRUE}. See also supporting material for further details. 

\subsection*{Inference from Phenotypic and Genetic Data jointly}
\begin{figure}[h]
\begin{tabular}{cc}
\hspace{-0cm}\includegraphics[width=8cm]{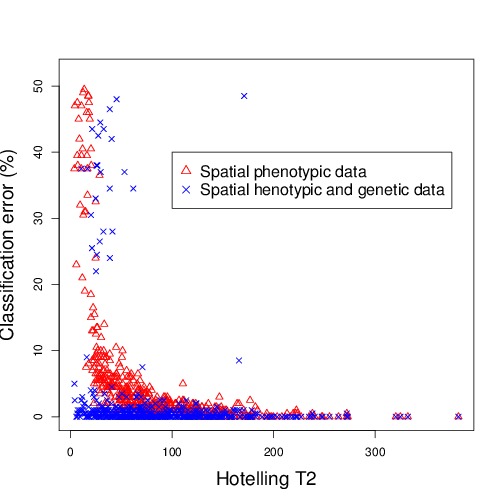} & 
\hspace{-0cm}\includegraphics[width=8cm]{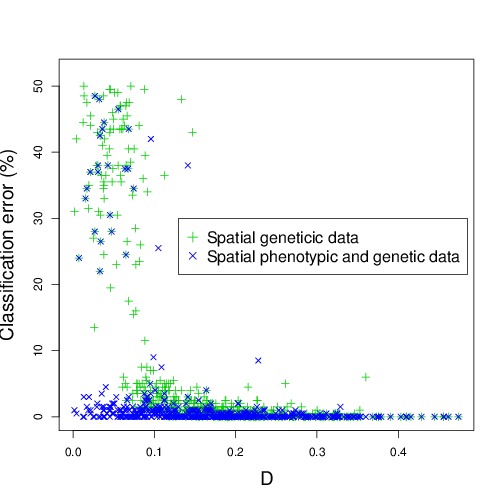} \\
\multicolumn{2}{c}{Average error: {\color{red} \footnotesize  $\boldsymbol  \triangle$} 5.2\%, {\color{green} \footnotesize  $\boldsymbol +$}8.7\%, 
{\color{blue} \footnotesize  $\boldsymbol \times$} 2.4\%}
\\
\end{tabular}
\caption{Classification error for 500 simulated data-sets consisting of 200 individuals belonging to $K=2$ clusters and recognized by $q=5$ quantitative variables and $l=10$ co-dominant loci. 
The variable plotted on the y-axis is the proportion of misclassified individuals using our method 
(after correction for potential label switching issues). 
}\label{fig:jointQG}
\end{figure}

We illustrate here how combining phenotypic and genetic data can improve the accuracy of inferences compared to inferences carried out from one type of data only. 
To do so, we simulated 500 data-sets consisting of two clusters each. There were five phenotypic variables and ten co-dominant genetic markers. 
We investigated a broad range of phenotypic and genetic differentiation and it appears that on average combining the two types of data increases 
the accuracy of inferences. See figure \ref{fig:jointQG}.\\

\section*{\sc Analysis of Data from the Literature}

\subsection*{Analysis of Iris  Morphometric Data}

\begin{figure}
\includegraphics[width=16cm,height=13cm]{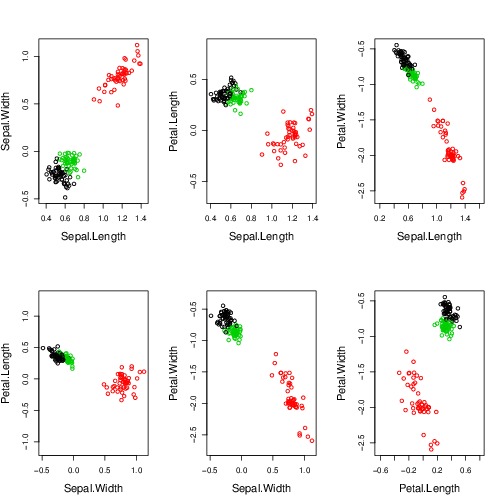}
\caption{Pairs plots of Fisher's Iris data (transformed into log shape ratios). Colors indicate individual species estimated by our method. 
The true number of species (three) is correctly estimated. Only 9  out of 150 individuals are misclassified.}\label{fig:Iris_gld}
\end{figure}

Fisher's  iris data-set \citep{Anderson35,Fisher36} gives the
     measurements in centimeters of the variables sepal length and
     width and petal length and width, respectively, for 50 flowers
     from each of 3 species of iris.  The species are {\em Iris setosa},
     {\em versicolor}, and {\em virginica}. 
We applied our method to the data transformed into log shape ratios \citep[see][and references therein]{Claude08}. 
Since the data are not georeferenced, we used the non-spatial prior. 
We launched ten independent MCMC runs. Seven of them return correctly $\hat{K}=3$, 
the other three runs return  $\hat{K}=4,5$ and $6$ respectively. 
Ranking the runs according to the average posterior density, 
the best run corresponds to one of the seven  runs that estimate $K$ correctly 
(according to the number of actual species in the data set).
This run achieves a classification error of $6$\% (see Fig. \ref{fig:Iris_gld}).
{\sc Mclust} returns an estimate of $K$ equal to 2 (raw data or log shape ratio data) 
and  50 out of 150 individuals are misclassified, 
thus failing to identify the three species of the data set. 

\subsection*{AFLP Data of {\em Calopogon} from Eastern North America and the Northern Caribbean}
The way our model deals with genetic data and the accuracy resulting from this method based on genetic data only has been investigated 
by \cite{Guillot05a,Guillot08a,Guillot08b,Guillot10b,Safner11,Guillot11b} and further discussion can be found in 
 \cite{Guillot09g}, however, to further illustrate the accuracy of our method when used with genetic data only, 
we study here a dataset produced and first analyzed by \citet{Goldman04}. 

This dataset consists of sixty {\em Calopogon} samples genotyped at 468 AFLP markers. 
\citet{Goldman04} identified the presence of five species ({\em C. barbatus, C. oklahomensis, C. tuberosus, C. pallidus, C. multiflorus}) 
and two hybrids specimens ({\em C. tuberosus $\times$ C. pallidus} and {\em C. pallidus  $\times$  C. multiflorus}).
According to \citet{Goldman04}, {\em C. tuberosus} has been widely considered to have three
varieties: var. {\em tuberosus}, var. {\em latifolius} and 
var. {\em simpsonii}. 
In addition, the dataset contains samples from two outgroups so that one could consider that 
the dataset contains up to eleven distinct species.  

We analysed this dataset under the same setting as the previous dataset. 
Under the UFM, the estimated $K$ ranges between 2 and 3 . The best run (in terms of average posterior density) corresponds to $\hat{K}=3$. 
In this clustering, one cluster contains the samples of the {\em  C. tuberosus} species, a second cluster merges the samples of the 
{\em  C. barbatus,  C. oklahomensis,  C. pallidus,  C. multiflorus} species and the hybrids. 
The last cluster contains the samples from the two outgroups.
Under the CFM, the estimated $K$ ranges between 7 and 8 . The best run (in terms of average posterior density) corresponds to $\hat{K}=8$. 
It  clusters the individuals of the various species as follows:
{\em  C. oklahomensis} / 
{\em   C. multiflorus} / 
{\em  C. barbatus} / 
{\em  C. pallidus}, {\em C. tuberosus $\times$ C. pallidus} and {\em C. pallidus  $\times$  C. multiflorus} / 
{\em  C. tuberosus tuberosus} except three samples /  
the three {\em  C. tuberosus tuberosus} previous  samples / two extra clusters for the outgroups.

\section*{\sc Analysis of  {\em Myodes}  Vole Data}

\subsection*{Data and statistical analysis }

We now study an original  dataset of geo-referenced genetic and phenotypic markers of the voles of the genus {\em Myodes } in Sweden. 
This dataset has several interests to investigate the efficiency of our method on a complex real case. (i) Fennoscandia has been 
recognised as a zone where the mitochondrial DNA of the northern red-backed vole {\em Myodes rutilus} introgressed its southern relative, 
the bank vole {\em M. glareolus} \citep{Tegelstrom87}. 
This makes the identification of these two species impossible based on common mitochondrial markers. 
(ii) The bank vole is further characterized by intra-specific lineages \citep{Deffontaine09}. 
Two of them are documented in Sweden \citep{Razzauti09}, providing a complex case for disentangling intra- and inter-specific structure. 
(iii) Both genetic and morphological data are available on this model to confront the structure provided by the two kinds of markers, and test for their combination.

The dataset consists of 182 individuals. These individuals were genotyped at 14 microsatellite loci \citep{Lehanse10}. 
The phenotypic dataset corresponds to a subsample of 69 individuals \citep{Ledevin10c}. 
We used measurements of the third upper molar shape, for which a phenotypic differentiation has been evidenced
at the phylogeographic scale \citep{Deffontaine09,Ledevin10}. 
The two-dimensional outline was manually registered from 
numerical pictures, starting from a comparable starting point among teeth \citep{Ledevin10}. For each molar, the outline is described by the 
Cartesian coordinates of 64 points sampled at equally spaced intervals along the outline.
These 64 landmarks are strongly correlated and therefore carry redundant information. 
To summarize this information into a lower number of variables and decrease the intensity of correlation between variables, 
we first performed an elliptic Fourier transform \citep[EFT][]{Kuhl82}. 
The EFT provides shape variables standardized by size, the Fourier coefficients that weight the successive functions of the EFT, 
namely the harmonics.
A study of the successive contribution of each harmonic to the description of the original outline showed that considering the 
first ten harmonics offered a good compromise between the number of variables and the efficient description of the outline \citep{Ledevin10}.
Then we performed a principal component analysis of the Fourier coefficients and retained the scores on the first five principal 
components, which contained more than 80\% of the variance (PC1=26.6\%, PC2=21.6\%, PC3=15.2\%, PC4=7.4\%, PC5=6.5\%).
These scores were used as phenotypic data input (the ${\bf y}$ data matrix) to our clustering method.

We analysed this dataset with our model first under the UFM allele frequency prior then under the CFM prior. 
For each allele frequency prior, we fed the model with five types of data combination:
using the georeferenced phenotypic data  under the spatial model (PS), 
using the phenotypic data  under the non-spatial model (PnS), 
using the georeferenced genetic data  under the spatial model (GS),
 using the genetic data  under the non-spatial model (GnS), 
using the georeferenced phenotypic and genetic data  under the spatial model (PGS).
In each case, we performed 10 independent MCMC runs of 100000 iterations discarding the first 10000 iterations as burnin.

\subsection*{Results}

For each type of analysis, we observed an excellent congruence across the ten independent MCMC runs. 
The UFM and the CFM model provide qualitatively similar results with a tendency of the CFM model to return slightly  larger estimates of $K$. 
While the CFM option has proven to detect finer differentiation than the UFM option (see analysis of AFLP data above), a detailed analysis 
and interpretation of the fine scale structure inferred by the CFM model would require extended data analysis, including some extra data 
still under production. We therefore focus on the results obtained under the UFM option.

\begin{figure}[h]
\thispagestyle{empty}
\vspace{-1.7cm}
\begin{tabular}{c}
\vspace{-1cm}\hspace{3cm}\textsf{Phenotypic \& Spatial - } $\mathsf{\hat{K}=2}$\\
\vspace{-.7cm}\hspace{3.5cm}\includegraphics[width=9.cm]{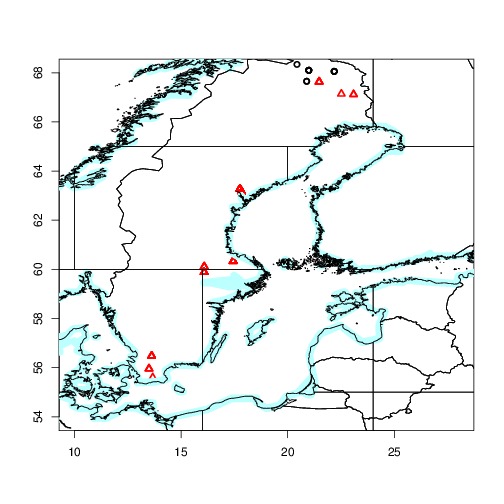}\\
\vspace{-.9cm}\hspace{3cm}\textsf{Phenotypic non Spatial -} $\mathsf{\hat{K}=2}$\\
\vspace{-.7cm}\hspace{3.5cm}\includegraphics[width=9.cm]{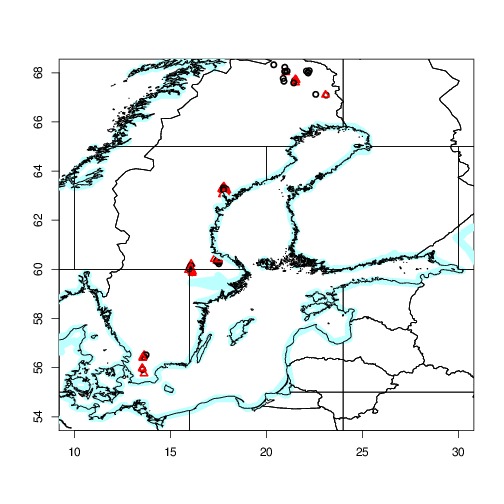}\\
\vspace{-.9cm}\hspace{3cm}\textsf{Genetic  \& Spatial -} $\mathsf{\hat{K}=4}$\\
\vspace{-.5cm}\hspace{3.5cm}\includegraphics[width=9.cm]{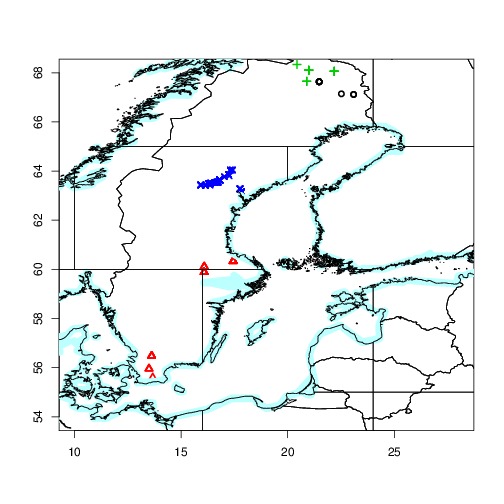}\\
\end{tabular}
\caption{Population structure inferred on the bank vole data. }
\label{fig:geno_UFM}
\end{figure}

In the analysis based on georeferenced  phenotypic data (PS), we inferred two clusters with one cluster in the top North of Sweden 
(Fig. \ref{fig:geno_UFM} top panel), 
all remaining samples belonging to the other cluster. 
These clusters correspond to the inter-specific differentiation between the red-backed vole to the North and the bank vole to the South.  
Analysing these data without spatial information (PnS), we also inferred also two clusters (Fig. \ref{fig:geno_UFM} middle panel).
The areas occupied by the two clusters under the PS and the PnS analyzes match in the sense that they both correspond to a top North vs. South dichotomy 
with a region of marked transition estimated to be along the same line in Swedish Lapland with a SW-NE orientation. 
In the PnS analysis, the clusters display  a large amount of spatial overlap with a regular North to South cline. 
In the analysis based on georeferenced genetic data (GS), we inferred the presence of four clusters. 
The most northern cluster corresponds to the samples identified as belonging to the top North cluster in the phenotypic clustering, 
and hence to the Northern red-backed vole (Fig. \ref{fig:geno_UFM} bottom panel). 
The three other clusters correspond to the intra-specific structure within the bank vole. 
This hierarchical pattern of inter- and intra-specific differences is confirmed by estimates of inter-population differentiation provided by Fst values. 
The top North population attributed to the red-backed vole appears as strongly differentiated from all other populations 
(N Sweden vs. NE Sweden: $\Fst$ = 0.15; N vs. Central Sweden: $\Fst$ = 0.19; N Sweden vs. South Sweden: $\Fst$ = 0.17). 
 In comparison, the differentiation is of smaller magnitude among bank vole populations (NE vs. C: $\Fst$ = 0.07; NE vs. S: $\Fst$ = 0.07; C vs. S: $\Fst$ = 0.06). 
Analysing these data without spatial information (GnS), we inferred four clusters whose locations match tightly those 
obtained under analysis GS (results not shown).
In the joint analysis of  georeferenced  phenotypic and genotypic data (PGS), we obtained results similar to those 
obtained  with georeferenced  genetic data  (results not shown).

\clearpage 
\section*{\sc Discussion}

\subsection*{Summary of Approach Proposed}

\subsubsection*{Main features}

We have proposed the first method to date for analyzing georeferenced  phenotypic and genetic data within a unified inferential framework, 
opening the way to combined analyses and robust comparison between markers. 
Our method takes as input any combination of phenotypic and genetic individual data and these data can 
be  optionally georeferenced. 
Analyses can be run on phenotypic and genetic data separately or jointly.
The main outputs of the method are  estimates of the number of homogeneous clusters and 
of cluster memberships of each individual. If analyses are made on georeferenced data, the method also 
provides an estimate of the spatial location of each cluster which can be displayed graphically 
in form of continuous maps (see program documentation for details on such graphic representation). 

Our approach is based on an explicit statistical model. This contrasts with model-free methods such as PAM which roughly speaking attempts 
to  cluster individuals in order to maximize some homogeneity criterion.
While such methods are fast and presumably robust to departure from specific model assumptions  
they are expected to behave poorly compared to methods based on an explict model that fits the data to a reasonable extent. 
This claim is supported by the recent study of \citet{Safner11} in the case of spatial genetic clustering methods.
In addition, because model-free methods do not rely on an explicit model, their output 
might be difficult to interpret or relate to biological processes. 

\subsubsection*{Main results from simulation study and analysis of classic data-sets}

\paragraph*{Inference from Phenotypic Data Only:}
All numerical results obtained here demonstrate the good accuracy of our method 
and its efficiency for identifying species and/or populations boundaries. 
It is excellent at avoiding false positives (i.e. at reporting $\hat{K}=1$ when $K$=1) 
and has a clear ability to reduce the error rate when the number of variables increases.
The method loses  accuracy  when it is given a difficult problem (i.e. when the true $K$ is large).
For a fixed number of iterations, it also has increasing difficulty to exploit fully all of the available information when the number of variables 
is large (cf. loss of accuracy for 50 variables compared to 20 variables), presumably due to loss of numerical efficiency in the MCMC algorithm.
We also noted that {\sc Mclust} is subject to similar difficulties for large number of clusters and/or large number of variables 
presumably due to the existence of multiple maxima of the likelihood.
In our method, this problem can be resolved to a certain extent by longer MCMC runs, an aspect not investigated in detail here.
Overall, our method offers a notable improvement over the non-spatial penalized maximum likelihood method of {\sc Mclust} used 
under its default set of options. 
One factor responsible for this improvement could be that our method exploits spatial information while  {\sc Mclust} does not. 
Results from section ``Analysis of classic data of the clustering literature'', where our method still provides better results 
than {\sc Mclust} even though the data are non-spatial, suggests this is not the sole factor. 
This might relate to model selection which is the second major difference between the two methods considered (Bayes vs. penalized maximum likelihood)
more so bearing that {\sc Mclust} considers a broad family of covariance structure while our method assumes conditional independence.

We also stress that the numerical values characterizing the accuracy of our method have to be taken with a grain of salt 
since  the model used to analyze the data matches exactly the model that generated them. 
This situation is a best case scenario and is unlikely to be  strictly met in real-life cases. 
However,  our results are informative about the potential of the method and evaluations of the iris data 
suggest a certain robustness of these results (see also analysis of crab morphometric data in supplementary material). 

As a final note, we warn the reader unfamiliar with clustering methods against overly pessimistic interpretation of  figure \ref{fig:Hotelling_nindiv=200_spatial=TRUE}.
From this figure, it seems that the methods lose accuracy very quickly as the 
the ``phenotypic differentiation'' decreases and are in general not so efficient. 
This is because detecting a hidden structure is a much harder statistical 
problem than testing the significance of a differentiation between two known clusters (the former involving many more parameters and hence uncertainty 
than the latter). 
More details are given in section "Power to test the significance of a known structure versus power to detect a hidden structure" 
of Supporting Material.

\paragraph*{Inference from Phenotypic and Genetic Data}

Genetic and phenotypic data can trace different evolutionary histories, for instance phylogenetic divergence for neutral genetic markers 
and adaptation for a morphological structure \citep{Renaud07,Adams09}.
Note that this is also true for any genetic marker that only traces its own evolutionary history in a phylogenetic dynamics 
\citep{Turmelle11}. Confronting the structure provided by different markers emerges more and more as a way to get 
a comprehensive view of the dynamics and processes of differentiation among and within species. Our method, by providing a 
unified inferential framework for analysing different kind of data, including phenotypic ones, appears as a significant 
improvement for valid confrontation between data sets. Furthermore, in situations when genetic and phenotypic patterns are suspected to coincide, 
making inference from genetic and phenotypic data jointly  has the potential to increase the power to detect 
boundaries between evolutionary units at different levels (populations, species).

\subsubsection*{Analysis of the {\em Calopogon}  AFLP data-set}
The ability of our model under the CFM prior
to detect and classify species is excellent.
This  dataset has been re-analyzed by \citet{Hausdorf10} who carried out a comparison of {\sc Structure}, 
{\sc Structurama}, a method known as ``field of recombination'' \citep{Doyle95} 
and a hybrid method mixing sequentially multidimensional scaling and model-based Gaussian clustering. 
The {\sc Structure} program and the ``field of recombination''method were not able to detect any structure. 
 {\sc Structurama} identified only three clusters and misclassifies 44\% of the samples. 
The hybrid method of \citet{Hausdorf10} identifies 5 clusters but misclassifies 15\% of the samples. 
Our method under the CFM prior also identifies 5 clusters but misclassifies only 5\% of the samples. 
Under the UFM model, the results we obtain are higly consistent with those obtained with the CFM.\\
We also refer the reader to the Supplementary Material where we analyze AFLP data of {\em Veronica (pentasepalae)} from the Iberian Peninsula and Morocco 
 produced  and first analysed by \citet{MartinezOrtega04}. 
The results we report  there confirm the excellent performance of our method compared to the four methods investigated by  \citet{Hausdorf10}. 
Finally, all the analysis carried out in the present article show that concerns of \citet{Hausdorf10} against methods for dominant markers 
based on Hardy-Weinberg equilibrium were not grounded, provided 
the dominant nature of AFLP markers is taken into account at the likelihood level as we did. 
We suspect that the poor performances of {\sc Structure} observed by \citet{Hausdorf10} relate to the procedure used 
to estimate $K$ \citep{Evanno05}, as noted earlier by \citet{Waples06}.

\subsubsection*{The {\em Myodes} data-set}
We confronted clustering hypotheses using various data subsets with or without spatial data and 
with or without genetic markers or morphometric variables. 
This shed new lights on the population structure of {\em Myodes}. 
The pattern of phenotypic and genetic differentiation can find an interpretation in a complex pattern of contact between species and populations.
The northernmost area corresponds to the narrow zone of possible overlap between  {\em Myodes glareolus} 
and its close northern relative {\em Myodes rutilus}. Both species are difficult to recognise based on external phenotypic characters, 
and impossible to identify based on common mitochondrial markers because of the introgression of {\em M. rutilus} mtDNA into the northern fringe 
of {\em M. glareolus} distribution.  
The northern cluster detected by our method corresponds most probably  to the occurrence of the northern red-backed vole {\em M. rutilus}, 
that tends to differ in molar shape from its relative {\em M. glareolus} \citep{Ledevin10b}.

The two analyses based on phenotypic data with and without spatial information  lead to slightly different results, 
the former suggesting the presence of an abrupt phenotypic discontinuity in the North while the latter suggests clinal variation 
(Fig. \ref{fig:geno_UFM} upper and middle panel). In absence of model fit criteria to assess the value of these two maps, we are reduced to speculate. 
We note however that these maps are congruent concerning the location of the main area of transition between the clusters and that the analysis based on spatial information 
is graphically more efficient at displaying the location of this transition. 
The bank vole molar shape has been shown to display a large variation even within populations, due to wear and developmental factors 
\citep{Guerecheau10,Ledevin10b}. This may render even clear cut inter-specific boundaries difficult to detect. 
Our georeferenced method may greatly help to make such signal emerge despite the intrinsic variability in the phenotypic markers. 
This suggests that our method could be viewed as an efficient 
generalisation of the methods aimed at detecting abrupt changes of \citet{Womble51} and \citet{Bocquet94}.

Regarding the additional clusters detected based on genetic data, the location of two of them suggests that they correspond to bank vole 
lineages already known in this region based on mitochondrial DNA data. Indeed, after the last ice age, Sweden has been 
recolonized by different populations separated  several hundreds of thousand years ago coming from the South and from the North of 
Fennoscandia \citep{Jaarola99,Razzauti09}. Our new data therefore confirm the existence of two different 
bank vole lineages in Sweden based on mitochondrial and now nuclear DNA markers. 
The existence of a fourth cluster located in Central Sweden strongly suggests that the contact zone 
between these two main lineages is situated in this latter region. Its origin may be attributed to hybridization 
between animals of the two genetic lineages. The discovery of this last cluster is new and it was never detected previously 
using only mitochondrial DNA marker. 

Combining phenotypic and genetic data in a joint analysis (PGS) did not allow us to detect any extra structure (map not shown), 
possibly because beyond the inter-specific phenotypic difference corresponding to the differentiation between top North and the rest of Sweden, 
a cline in molar shape exists through Sweden that is roughly congruent with the genetic clusters (data not shown).
It shows that the confrontation between data sets may be as informative as a joint analysis, by providing clues about 
the hierarchical pattern of differentiation. 
Morphometric clusters evidenced here inter-specific differences between red-backed and 
bank voles whereas based on microsatellite data, both inter- and intra-specific levels of differentiation emerged as separate clusters. 
The structure of genetic differentiation corroborates this interpretation. The inter-specific differentiation of the top North cluster 
from the rest of Sweden is indeed much stronger than the intra-specific differentiation among the bank vole populations from North-East, 
Central and South Sweden.  
Combining together both data types allows us to interpret the complex phylogeographic structure of this species and helps to distinguish 
differences between true species and populations within a species.

\subsection*{Future Extensions}

Our method is based on an assumption of independence of the phenotypic variables within each cluster. 
This does not amount to independence between these variables globally. Indeed, the fact that phenotypic variables are sampled with cluster-specific 
parameters does include a correlation (similarly to the dependence structure assumed in a linear mixed model). However our method does not 
deal with residual dependence not accounted for at the cluster level such as that generated by allometry.
Results from simulations and classic datasets suggest that this can be partially dealt with by pre-processing the data 
(e.g. transforming raw data into log-shape ratio). 
Several other  
procedures may be applied for avoiding or reducing problems with  
covariation among phenotypic variable. For example, working on  
principal components rather than on raw data may help in this task.  
Procedures such as the Burnaby approach \citep{Burnaby66} may also allow  
to remove covariance structures due only to growth or other  
confounding factors that the user may wish to filter out. 
A more rigorous approach would consist in allowing the variables to covary within clusters 
which would also allow one to quantify these covariations.


\subsection*{Potential Applications}


Evolutionary biology has been flooded by molecular data in the recent years. However, efficient methods to deal with phenotypic data alone 
are still needed when this type of data is the only available. 
This includes the important case of fossil data. We note that in systematic paleontology, 
the methods used are often simpler than those discussed in the present paper and chosen as a matter of tradition in the field 
rather than on objective basis. Implementing our method in a free and  user-friendly program should help provide 
more objective methods in this context.

Our method was specifically tailored for biometric/morphometric measurements which are typically obtained at a few tens 
of phenotypic variables. The method proposed is therefore computer intensive and not expected to be well suited for large datasets
such  as expression data produced in functional genomics. However, in the situations where  the scientist is able to select some variables 
of particular interest and reduce the dimensionality of the model (as we did for our analysis of the {\em Myodes} molar shape data), 
our method could be used and play a role in the emerging field of landscape genomics \citep{Schwartz10}. 


The sub-model for genetic data used here was presented and discussed in detail by \citet{Guillot05a} and  \citet{Guillot08b}.
It has been used mostly to analyse variation and structure in neutral nuclear markers \citep{Guillot09g} and proved useful 
to detect and quantify fine-scale structure typical of landscape genetics studies. 
The novel possibility brought here to combine it with morphometrics data might popularize this genetic model among scientists 
interested in larger spatial and temporal scale typical of phylogeography. 
In the latter field, the use of mtDNA is common. As noted earlier, the analysis of such non-recombining DNA sequence data  
using our method is technically possible and meaningful by recoding the various observed haplotypes as different alleles of the same locus. 
We stress that this approach is an expedient which incurs a considerable loss of information and that our approach 
should not be viewed as a substitute to those that model the genealogy of genes (including the mutational process) explicitly. 
Extending our model to deal with non recombining DNA in a more rigorous way is a natural direction for future work. 

Our method for the combined analysis of phenotypic and genetic data can be used to assess 
the relative importance of random genetic drift and directional natural
selection as causes of population differentiation in quantitative traits, and
to assess whether the degree of divergence in neutral marker loci predicts the degree of
divergence in quantitative traits \citep{Merila01}. 
Furthermore, our method should be  useful in the study of hybrid zones where, as noted by \citet{Gay08}, 
comparing clines of neutral genetic markers with clines of traits known to be under selection also indicates the
extent to which the overall genome is under selection.

Lastly, because phenotypic and genetic markers may reflect different evolutionary or demographic history, 
combined analyses can help to understand the hierarchy between evolutionary units (species and populations) 
as shown in the {\em Myodes} example.

\newpage

\hspace{1cm}
\paragraph*{\sc Computer program availability:}
The model presented here will be available soon as part of a new version of the R package {\sc Geneland} (version $\geq$ 4.0.0). 
Information will be found on the program homepage {\tt http://www2.imm.dtu.dk/\urltilda gigu/Geneland/}.
\hspace{.5cm}
\paragraph*{\sc Acknowledgments:}
The first author is most grateful to Cino Pertoldi for discussions that prompted him to develop the model for morphometric data.
Our work benefitted from discussions with Jean-Marie Cornuet and comments of Andrew J. Crawford.
Part of the original data of the {\em Myodes} analysis belong to Bernard Lehanse's Master thesis (genetic data). 
We thank him for sharing these data with us. 
We are also grateful to Montse Mart{\'i}nez-Ortega and Doug Goldman for making there  data available to us.  
This work has been supported  by the French National Research Agency 
(project EMILE, grant ANR-09-BLAN-0145-01) and the Danish Centre for Scientific Computing (grant 2010-06-04).

\clearpage



\clearpage
\appendix

\section*{\sc Appendix: Detail of MCMC Inference Algorithm}

\subsection*{Overview}

The vector of unknown parameters is 
${\boldsymbol \theta}=(K,\lambda,m,{\bf u},{\bf c},{\bf f},{\bf \tilde{f}},{\bf d},
{\boldsymbol \mu},{\boldsymbol \sigma},{\boldsymbol \beta})$ 
which can be decomposed into  ${\boldsymbol \theta}_S=(\lambda,m,{\bf u},{\bf c})$, ${\boldsymbol \theta}_G=({\bf f},{\bf \tilde{f}},{\bf d})$ 
and ${\boldsymbol \theta}_M= ({\boldsymbol \mu},{\boldsymbol \sigma},{\boldsymbol \beta})$ 
blocks of parameters of the spatial, genetic and phenotypic data respectively.
We alternate block updates of Metropolis-Hastings or Gibbs type and also trans-dimensional updates 
involving changes of $K$ and of parts of other parameters. The updates of blocks of parameters that do not involve 
phenotypic data are described in \cite{Guillot05a} and \cite{Guillot08b}. We describe below updates involving phenotypic data. 

\subsection*{Joint Updates of $({{\bf c}},{\boldsymbol \mu}, {\boldsymbol \sigma}$)}
We update jointly  ${{\bf c}},{\boldsymbol \mu}$ and ${\boldsymbol \sigma}$ as follows.
We propose a new vector  ${\bf c}^*$ by picking two clusters at random and re-assigning some individuals 
of one of those two clusters to the other one at random.  Then we propose $\boldsymbol \mu$ and $\boldsymbol \sigma$ 
by sampling from the full conditional distribution $\pi({\boldsymbol \mu},1/{\boldsymbol \sigma}^2|{{\bf y}, {\bf c}^*})$.
The Metropolis-Hastings ratio is
\begin{eqnarray}
R & = & \frac{\pi({\boldsymbol \theta}^* | {\bf y})}{\pi({\boldsymbol \theta} | {\bf y})} %
\frac{q({\boldsymbol \theta}|{\boldsymbol \theta}^*)}{q({\boldsymbol \theta}^*|{\boldsymbol \theta})} \nonumber \\
  & = &  \frac{\pi({\boldsymbol \mu}^*,{1/{\boldsymbol \sigma^2}}^*, {\bf c}^* | {\bf y})}%
{\pi({\boldsymbol \mu},{1/{\boldsymbol \sigma^2}}, {\bf c} | {\bf y})}%
 \frac{q({\boldsymbol \mu},{1/{\boldsymbol \sigma^2}}| {\bf c})}%
{q({\boldsymbol \mu}^*,{1/{\boldsymbol \sigma^2}}^*| {\bf c}^*)}
\frac{q({\bf c}|{\bf c}^*)}{q({\bf c}^*|{\bf c})} \nonumber \\
 & = & \frac{\pi({\bf c}^* | {\bf y})}{\pi({\bf c} | {\bf y})}%
 \frac{\pi({\boldsymbol \mu}^*,{1/{\boldsymbol \sigma^2}}^*| {\bf c}^* , {\bf y})}%
{\pi({\boldsymbol \mu},{1/{\boldsymbol \sigma^2}}| {\bf c} , {\bf y})}%
 \frac{\pi({\boldsymbol \mu},{1/{\boldsymbol \sigma^2}}| {\bf c} , {\bf y})}%
{\pi({\boldsymbol \mu}^*,{1/{\boldsymbol \sigma^2}}^*| {\bf c}^* , {\bf y})}%
\frac{q({\bf c}|{\bf c}^*)}{q({\bf c}^*|{\bf c})} \nonumber \\
 & = &\frac{\pi({\bf c}^* | {\bf y})}{\pi({\bf c} | {\bf y})}%
\frac{q({\bf c}|{\bf c}^*)}{q({\bf c}^*|{\bf c})}  \label{eq:update_c_mu_sigma_simple}
\end{eqnarray}

Interestingly, the latter expression does not depend on $({\boldsymbol \mu}^*,{{\boldsymbol \sigma^2}}^*)$, which in principle would allow 
us to decide whet her a new state ${\boldsymbol \theta}^*$ is accepted  prior to proposing $({\boldsymbol \mu}^*,{{\boldsymbol \sigma^2}}^*)$. 
Unfortunately, expression (\ref{eq:update_c_mu_sigma_simple}) can not be used as $\pi({\bf c} | {\bf y})$ is not known analytically under the present model. 
 The ratio in equation (\ref{eq:update_c_mu_sigma_simple}) has therefore to be written as 
\begin{eqnarray}
R & = & 
\frac{\pi({\bf y} | {\boldsymbol \mu}^*,{1/{\boldsymbol \sigma^2}}^*, {\bf c}^*)}%
{\pi({\bf y} | {\boldsymbol \mu},{1/{\boldsymbol \sigma^2}}, {\bf c})}%
\frac{\pi({\bf c}^*)}{\pi({\bf c})}%
\frac{\pi({\boldsymbol \mu}^*,{1/{\boldsymbol \sigma^2}}^*)}%
{\pi({\boldsymbol \mu},{1/{\boldsymbol \sigma^2}})}%
\frac{\pi({\boldsymbol \mu},{1/{\boldsymbol \sigma^2}} |  {\bf y} , {\bf c})}%
{\pi({\boldsymbol \mu}^*,{1/{\boldsymbol \sigma^2}}^* |  {\bf y} , {\bf c})}%
\frac{q({\bf c}|{\bf c}^*)}{q({\bf c}^*|{\bf c})}
 \label{eq:update_c_mu_sigma_notsimple}
\end{eqnarray}
which involves only analytically known expressions. 

\subsection*{Joint Updates of $(K,{{\bf c}},{\boldsymbol \mu}, {\boldsymbol \sigma}$)}
We take  the same strategy as \citet{Guillot05a}.
 The algorithm follows ideas of \citet{Richardson97}. 
It consists in updating $K$ by proposing to split a cluster into two clusters or merge two clusters, 
in a way that complies with the spatial constraints and multivariate nature of  the model. 
Since we use the natural prior conjugate family for parameters ${\boldsymbol \mu}^*$ and ${\boldsymbol \sigma}^*$ 
the full conditional $\pi({\boldsymbol \mu}, 1/{\boldsymbol \sigma^2}^* | {\bf y},K^*,{\bf c}^*)$ 
is available and can be used as proposal distribution as advocated for example by \citet{Godsill01}. 
The acceptance ratio takes essentially the same form as in equation \ref{eq:update_c_mu_sigma_notsimple} although it is 
now a genuine transdimensionnal move.

\subsection*{Detail on Hyper-Parameters}
Although we do not use exactly the same prior structure as \citet{Richardson97}, we follow largely these authors.
We take $\xi_j = \sum_i y_{ij}$, $h_j=\kappa_j=2/R_j^2$ where $R_j$ is the range of observed values of the $j$-th phenotypic variable.
$\beta_j | g_j ,h_j, \sim {\cal G}(g_j,h_j)$. 
We also set $\alpha_j =2$ and $g_j=1/2$.
Since $E[1/\sigma^2] = \alpha/\beta$,  $\beta$ represents $2/E[1/\sigma^2]$. 
Also $1/2h$ represents the prior mean of beta.

\end{document}